\PassOptionsToPackage{table,xcdraw}{xcolor}
\documentclass[letterpaper]{article} 
\usepackage{aaai2026}  
\usepackage{times}  
\usepackage{helvet}  
\usepackage{courier}  
\usepackage[hyphens]{url}  
\usepackage[colorlinks=true]{hyperref}
\hypersetup{
    citecolor=blue!70!black,    
    linkcolor=green!50!black,     
    urlcolor=magenta   
}
\usepackage{graphicx} 
\usepackage{natbib}  
\usepackage{caption} 
\usepackage{algorithm}
\usepackage{algorithmic}
\usepackage{enumitem}

\usepackage{newfloat}
\usepackage{listings}
\usepackage{xspace}
\usepackage{amsmath}
\usepackage{amssymb}
\usepackage{booktabs}
\usepackage{subcaption} 
\usepackage{adjustbox}
\DeclareCaptionStyle{ruled}{labelfont=normalfont,labelsep=colon,strut=off} 
\floatstyle{ruled}
\newfloat{listing}{tb}{lst}{}
\floatname{listing}{Listing}
\newcommand{\method}{MetaDiT\xspace}
\title{\method: Enabling Fine-grained Constraints in High-degree-of Freedom Metasurface Design}
\author {
   Hao Li\textsuperscript{\rm 1,}\thanks{haolicq.ai.research@gmail.com},
    Andrey Bogdnaov \textsuperscript{\rm 1,2,}\thanks{a.bogdanov@hrbeu.edu.cn}
}
\affiliations {
    \textsuperscript{\rm 1}Qingdao Innovation and Development Center, Harbin Engineering University, Qingdao 266000, Shandong, China\\
    \textsuperscript{\rm 2}School of Physics and Engineering, ITMO University, St. Petersburg 197101, Russia\\
    $^*$haolicq.ai.research@gmail.com, $^\dagger$a.bogdanov@hrbeu.edu.cn
}

\usepackage{bibentry}

\begin{document}

\maketitle

\begin{abstract}
Metasurfaces are ultrathin, engineered materials composed of nanostructures that manipulate light in ways unattainable by natural materials. Recent advances have leveraged computational optimization, machine learning, and deep learning to automate their design. However, existing approaches exhibit two fundamental limitations: (1) they often restrict the model to generating only a subset of design parameters, and (2) they rely on heavily downsampled spectral targets, which compromises both the novelty and accuracy of the resulting structures. The core challenge lies in developing a generative model capable of exploring a large, unconstrained design space while precisely capturing the intricate physical relationships between material parameters and their high-resolution spectral responses. In this paper, we introduce \method, a novel framework for high-fidelity metasurface design that addresses these limitations. Our approach leverages a robust spectrum encoder pretrained with contrastive learning, providing strong conditional guidance to a Diffusion Transformer-based backbone. Experiments demonstrate that \method outperforms existing baselines in spectral accuracy, we further validate our method through extensive ablation studies. Our code and model weights will be open-sourced to facilitate future research.
\end{abstract}

\begin{links}
    \link{Code}{https://github.com/JessePrince/metadit.git}
\end{links}

\section{Introduction}
Metasurfaces are ultrathin, engineered materials composed of nanostructures that manipulate light in ways natural materials cannot~\cite{jeong2024review,koshelev2023bound,koshelev2018asymmetric}. Unlike bulky traditional optics (e.g., lenses), metasurfaces achieve precise wave control at subwavelength scales~\cite{kildishev2013planar,khorasaninejad2017metalenses}, enabling applications like ultracompact cameras~\cite{kim2024metasurface,park2024all}, AR/VR displays~\cite{aththanayake2025tunable,tian2025achromatic}, communications~\cite{xu2025dual,fu2025fundamentals} and optical computing~\cite{zhou2024optical,hu2024diffractive}. However, designing these materials is challenging due to their high-dimensional parameter space. Traditional approaches rely heavily on human intuition and iterative trial-and-error, which are often inefficient and suboptimal. To overcome these limitations, researchers have turned to inverse design methods: leveraging computational optimization~\cite{wang2023inverse,li2023inverse} and machine learning~\cite{al2023gaussian,tian2024high,chen2025generative} to discover metasurface structures that achieve target electromagnetic (EM) responses. These techniques not only accelerate the design process but also enable the discovery of novel configurations. Inverse design thus serves as a critical pathway toward scalable and optimized metasurface engineering.

Recent advances in artificial intelligence have spurred significant interest in applying deep learning models to inverse design problems~\cite{tanriover2022deep,yang2025exploring,dong2025advanced,saifullah2025deep}. Over the past decade, we have witnessed remarkable progress in this area, particularly with the emergence of generative models such as generative adversarial networks (GANs)~\cite{goodfellow2014generative,liu2018generative,so2019designing,yeung2021global}, variational autoencoders (VAEs)~\cite{kingma2013auto,tran2025deep,kojima2023inverse}, and diffusion models~\cite{ho2020denoising,niu2023diffusion,Zhangdiffusionmetasurface,zhang2024addressing}. These approaches have enabled the generation of multiple solutions for novel material designs that meet desired optical behaviors, proving to be a powerful engine for creating novel metasurface materials.

\begin{figure}[t]
    \centering
    \includegraphics[width=1\linewidth]{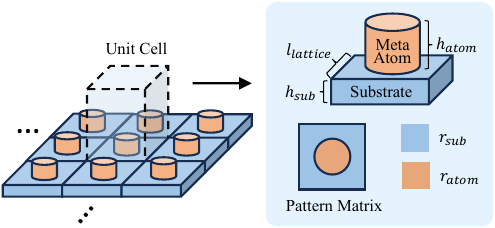}
    \caption{Illustration of metasurface material and the unit cell. The color represents different refractive index.}
    \label{fig:metaatom}
\end{figure}

Designing metasurfaces are designing their constituent materials, geometric configurations, and structural parameters. While recent advances have leveraged powerful generative models for metasurface design, we identify two critical limitations in the current paradigm. First, prior works~\cite{Zhangdiffusionmetasurface,zhang2024addressing,seo2025physics,niu2023diffusion} typically formulate the design task as a conditional generation problem, wherein key attributes, such as meta-atom thickness and lattice constants are manually fixed. This effectively reduces the model's role to mapping a target EM response to a geometry within a constrained, pre-selected subspace, thereby limiting the diversity and novelty of generated designs. Second, these approaches often rely on heavily downsampled target spectra (e.g. 12x downsampled in~\cite{Zhangdiffusionmetasurface}), which simplifies the optimization objective but compromises the physical fidelity of the resulting structures. Although low reconstruction error may be achieved on the coarse target, the final designs can exhibit undesirable frequency-dependent behaviors. We argue that the core challenge lies in developing a high-capacity generative model that can operate over a large, unconstrained design space while faithfully capturing the \textbf{underlying physical relationships} between material parameters and their high-resolution spectral responses, as governed by Maxwell’s equations. This leads us to the central question of our work:

\begin{center}
    \emph{How can we develop a generative framework that simultaneously optimizes all design parameters while precisely satisfying high-resolution spectral constraints?}
\end{center}

In this paper, we propose \textbf{\method}, a novel generative framework to address these limitations. We first develop a spectrum encoder equipped with sequence attention and channel attention mechanisms to extract rich semantic features from input spectra. To align spectral and structural representations, the encoder is trained using a contrastive learning objective alongside a Vision Transformer (ViT)~\cite{dosovitskiy2020image} that processes material geometries. Subsequently, we encode the metasurface material design as an image with three channels, and the pretrained spectrum encoder is leveraged to guide a Diffusion Transformer (DiT) based~\cite{peebles2023scalable} diffusion model for material generation. To enable fine-grained conditioning, we introduce a coarse-to-fine conditioning scheme: the coarse spectral embedding is injected via adaLN~\cite{perez2018film}, while the fine-grained embedding is concatenated with image tokens, facilitating in-context learning through self-attention. Furthermore, we employ the Accumulated Absolute Error (AAE) to capture localized failures that may be overlooked by conventional averaging metrics. We also introduce the AAE\&K metric, defined as the maximum AAE over $K$ independently generated designs for a given target spectrum, to assess the model’s consistency in producing diverse yet accurate solutions.

Experimental results show that \method significantly outperforms existing methods in designing novel metasurface materials with all variable parameters in dataset and fine-grained spectral targets. Specifically, \method reduces MAE and AAE by \textbf{52.2\%} compared to a vanilla DiT baseline, and surpasses MetaDiff~\cite{Zhangdiffusionmetasurface}, a model specifically designed for metasurface generation, by \textbf{39.1\%}. We conduct comprehensive ablation studies to assess the impact of each component in \method, confirming the effectiveness of our architectural design and training strategy. Moreover, we explore the scalability of \method: Can we obtain better performance by increasing model capacity? 

Our contributions can be summarized as follows:
\begin{itemize}
    \item We propose \method, a novel method that generates design parameters under high-resolution spectral constraints, with the flexibility to optimize all parameters when available.
    \item We propose novel metrics and perform extensive experiments to evaluate \method's performance and systematically analyze the impact of each design component.
    \item We open-source our code and all model weights in the hope of paving the way for the community to develop more powerful models.
\end{itemize}

\section{Related Works}
\subsection{Diffusion Generative Models}
\label{sec:diffgen}
Diffusion models~\cite{ho2020denoising} have emerged as highly capable generative frameworks, driving significant advances in image~\cite{dhariwal2021diffusion,nichol2021glide,rombach2022high,tumanyan2023plug} and video generation~\cite{blattmann2023stable,guo2023animatediff,tu2024motioneditor,wang2024magicvideo}. While early diffusion models primarily relied on U-Net~\cite{ronneberger2015u} backbones, the Diffusion Transformer (DiT)~\cite{peebles2023scalable} has demonstrated superior training stability and scalability, becoming the dominant architecture in many recent works~\cite{esser2024scaling,brooks2024_video_generation_world_simulators,kong2024hunyuanvideo}. Owing to their strong generative capacity, diffusion models have also been adopted in scientific domains, including molecular~\cite{wang2025diffspectra,liu2025next} and material generation~\cite{xie2021crystal,zeni2023mattergen}. Recently, researchers have begun applying diffusion models to metasurface design~\cite{Zhangdiffusionmetasurface,zhang2024addressing,seo2025physics,niu2023diffusion}, demonstrating their potential to generate novel, high-performance structures. However, existing approaches typically generate only a subset of design parameters and rely on downsampled, coarse spectral constraints, simplifying the problem but ultimately limiting the novelty and physical fidelity of the generated materials.

\subsection{Inverse Design of Metasurfaces}
Designing metasurfaces entails selecting constituent materials, configuring geometric layouts, and tuning structural parameters to achieve desired EM behavior across a frequency range. Early approaches employed computational optimization techniques~\cite{wang2023inverse,li2023inverse}, which, while provably convergent, require costly forward–adjoint field evaluations and struggle to scale in high-dimensional design spaces. To address these limitations, researchers have explored machine learning-based methods, framing inverse design as a conditional generative task. GANs\cite{liu2018generative,so2019designing,yeung2021global} and VAEs\cite{kingma2013auto,tran2025deep,kojima2023inverse} can generate diverse candidate structures in a single forward pass. However, GANs suffer from training instability, while VAEs often produce blurry reconstructions that compromise spectral fidelity. More recently, diffusion models have emerged as state-of-the-art in metasurface material design. Works such as~\cite{Zhangdiffusionmetasurface,zhang2024addressing,seo2025physics} have demonstrated that diffusion-based frameworks can outperform GAN and VAE baselines. Nevertheless, as discussed in Section~\ref{sec:diffgen}, existing models simplify the design task by restricting parameter coverage and using coarse spectral constraints. To address these limitations, we propose \method, which integrates a contrastively pretrained spectrum encoder with a Diffusion Transformer based backbone. \method enables fine-grained control and exploration of a more complete metasurface design space.

\section{Preliminaries}
\subsection{Metasurfaces and Scattering Spectrum}
Metasurfaces are planar arrays of subwavelength dielectric structures, where each periodic unit cell $U$ is defined by a set of geometric and material parameters. As shown in Figure~\ref{fig:metaatom}, a unit cell consists of two key components:
\begin{enumerate}
    \item A \textbf{substrate} with refractive index $r_{\text{sub}}\in\mathbb{R}$ and thickness $h_{\text{sub}}\in\mathbb{R}$.
    \item A \textbf{meta-atom} with refractive index $r_{\text{atom}}\in\mathbb{R}$, thickness $h_{\text{atom}}\in\mathbb{R}$, and a binary geometric pattern matrix $P\in \mathbb{F}_2^{n\times n}$ encoding its structure (where $n$ is the spatial resolution of the pattern).
\end{enumerate}

The lattice constant $l_{\text{lattice}}\in \mathbb{R}$ governs the periodicity of the array. Together, these parameters fully describe the metasurface’s optical properties.

The optical responses of a metasurface is characterized by its \textbf{scattering spectrum}, a complex-valued function $S(f)\in \mathbb{C}$ that describes the amplitude and the phase of scattered light at frequency $f$. This spectrum is goverened by the aforementioned material parameters, which collectively determine resonant scattering behavior.

\subsection{Problem Definition}
The goal of metasurface inverse design is to automatically generate a unit cell $U$ that achieves a desired scattering spectrum $S(f)$. Traditional approaches rely on human expertise and iterative trial-and-error, which are often computationally expensive. Recent advances leverage leraned models $M$ to directly predict unit cell geometries $\hat U = M(S)$ from target spectra. The generated designs are then validated via electromagnetic simulations, through which we can calculate the spectrum of a metasurface material. For convenience, we also refer $U$ as the material design or structure.

\section{Method}
\label{sec:method}

\subsection{Dataset and Encoding}
\label{sec:data encode}
\begin{figure}[t]
    \centering
    \begin{minipage}[t]{0.48\linewidth}
        \centering
        \includegraphics[width=\linewidth]{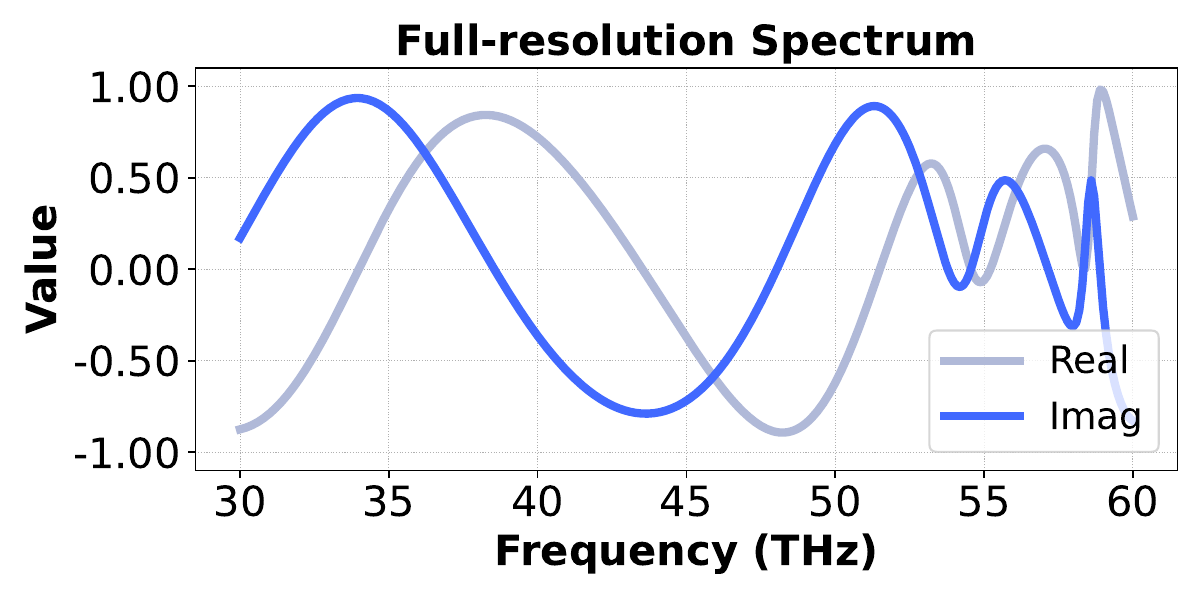}\\
        \includegraphics[width=\linewidth]{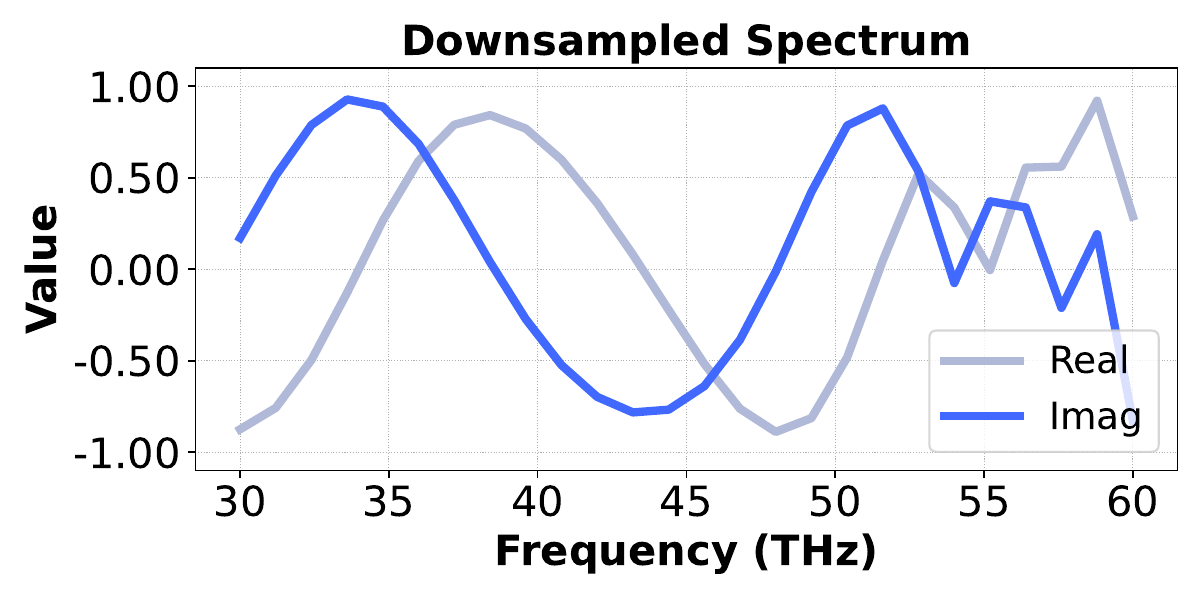}
    \end{minipage}%
    \hspace{0.02\linewidth}
    \begin{minipage}[t]{0.48\linewidth}
        \centering
        \adjustbox{valign=m}{\includegraphics[width=0.9\linewidth]{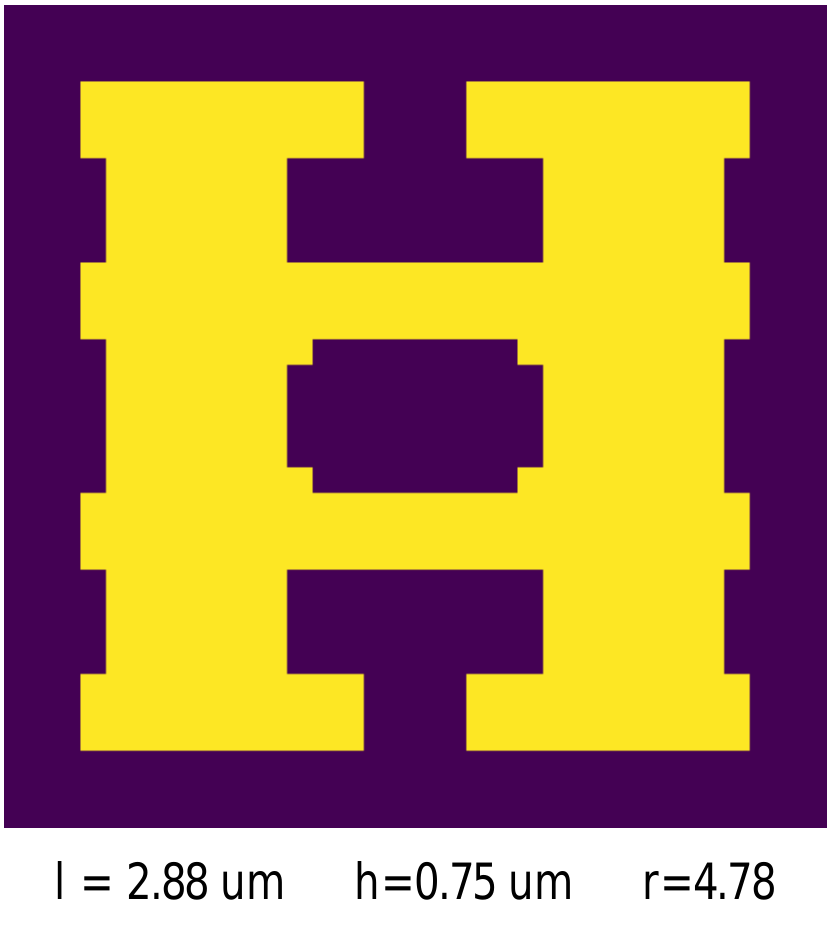}}
    \end{minipage}
    \caption{\textbf{An example of a metasurface material sample.} (Top left) Full-resolution scattering spectrum. (Bottom left) Downsampled spectrum, where the peak near $\sim$58~THz is attenuated due to loss of resolution. (Right) Corresponding pattern matrix; yellow regions (ones) indicate the meta-atoms, while dark purple (zeros) denotes the substrate. The $l$, $h$, $r$ represents $l_{\text{lattice}}$, $h_{\text{atom}}$ and $r_{\text{atom}}$ respsectively.}
    \label{fig:meta example}
\end{figure}

Following~\cite{Zhangdiffusionmetasurface}, we adopt the dataset introduced in~\cite{an2020deep}, which contains 170k+ metasurface designs with a high degree of geometric variability generated via a randomized algorithm. Each metasurface unit cell is represented by a binary pattern matrix of size $64 \times 64$, along with three continuous design parameters: the atom refractive index $r_{\text{atom}}$, atom thickness $h_{\text{atom}}$, and lattice constant $l_{\text{lattice}}$. The substrate refractive index $r_{\text{sub}}=1.4$ and height $h_{\text{sub}}=2\mu m$ are held constant throughout the dataset.

We encode each unit cell as a three-channel image $U \in \mathbb{R}^{3 \times 64 \times 64}$, where the three channels respectively represent the design parameters $r_{\text{atom}}$, $h_{\text{atom}}$, and $l_{\text{lattice}}$. Within each channel, the original binary pattern matrix is preserved in structure: positions with value one are replaced by the corresponding design parameter, while zeros remain unchanged. This encoding scheme enables the model to generate all structural parameters jointly, in contrast to prior works that treat $r_{\text{atom}}$, $h_{\text{atom}}$, and $l_{\text{lattice}}$ as fixed conditioning inputs.

The target in the dataset is the transmission scattering spectrum. It is provided at 301 discrete frequency points and is encoded as a two-channel sequence $S \in \mathbb{R}^{301\times 2}$, where the channels represent real and imaginary values. Unlike prior works~\cite{Zhangdiffusionmetasurface,zhang2024addressing}, which downsample the spectrum to simplify the target, we preserve the full resolution to enable the model to capture and satisfy fine-grained spectral constraints. An example of a material structure and its corresponding spectrum is shown in Figure~\ref{fig:meta example}. Notably, the downsampled spectrum exhibits attenuation of the peak near $\sim$58~THz, highlighting the loss of fine-grained spectral information due to resolution reduction.

\begin{figure*}[t]
    \centering
    \includegraphics[width=0.9\linewidth]{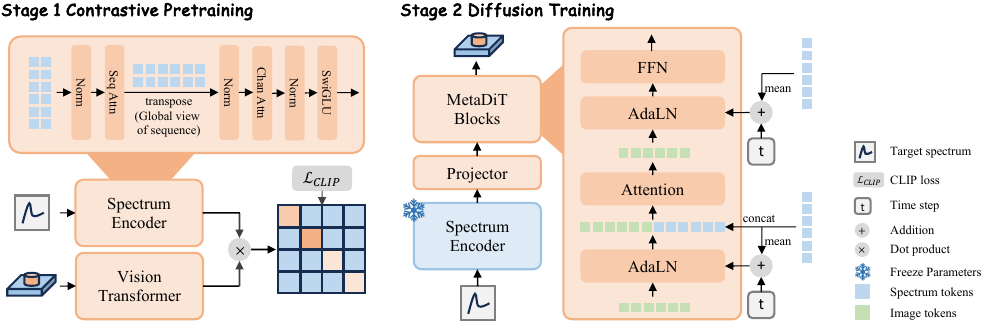}
    \caption{\textbf{Overview of \method}. We first train a spectrum encoder using contrastive learning, then the spectrum feature is pre-computed and fed into \method backbone for diffusion training.}
    \label{fig:main diagram}
\end{figure*}

\subsection{Architecture of \method}
The core objective of \method is to learn the intrinsic relationship between a material’s structure and its corresponding spectral response. To this end, we design the key components of \method by addressing two central questions that guide our architectural and training decisions.
\subsubsection{1. How can we effectively encode fine-grained spectral conditions?}
Prior works~\cite{Zhangdiffusionmetasurface,zhang2024addressing} encode the spectrum as a single global feature vector for conditioning (i.e. $D$-dimensional vector and is projected to the model hidden size), which we argue overlooks the semantic structure inherent in spectral data. Preserving the sequence format allows the model to attend to localized spectral patterns that are critical for accurate material generation.

We propose to use a Transformer-based~\cite{vaswani2017attention} encoder to encode the spectrum in its sequence format. To enable effective representation learning within the Transformer encoder, we first project the input spectrum from its raw form $S\in \mathbb{R}^{301\times 2}$ into a higher-dimensional feature $X_S\in \mathbb{R}^{301\times D_S}$ using a learnable linear embedding layer, where $D_S$ is the hidden size of the spectrum encoder. We then apply position embedding and stack $L$ layers of Transformer-based encoder to extract the feature.

However, we observe that the scattering spectrum is inherently a dual-variate sequence, where the embedded representation encodes both amplitude and phase components that are intrinsically coupled by physical laws. This coupling introduces rich inter-channel dependencies that are crucial for accurate modeling. While in typical Transformer encoder, the self-attention applied along the frequency (sequence) axis captures long-range dependencies and dispersion patterns, it alone is insufficient to fully represent the complex-valued nature of the spectrum. To address this limitation, we incorporate an additional attention mechanism along the channel axis, enabling the model to globally attend over the entire spectrum and explicitly model the inter-channel relationships. This dual-attention design facilitates more effective feature re-weighting and enhances the model's capacity to capture the underlying physical structure of the data. Our encoder block is formulated as follows:
\begin{equation}
\begin{aligned}
    &X_l^{(1)} = X_l + \text{Attn}_\text{seq}(\text{Norm}(X_l)) \\
    &X_l^{(2)} = X_l^{(1)} + \left[\text{Attn}_\text{chan}(\text{Norm}((X_l^{(1)})^\mathsf{T}))\right]^\mathsf{T} \\
    &X_{l+1} = X_l^{(2)} + \text{FFN}(\text{Norm}(X_l^{(2)}))
\end{aligned}
\end{equation}
Here, $X_l$ denotes the input feature at layer $l$, $\text{Attn}_\text{seq}$ and $\text{Attn}_\text{chan}$ represent self-attention along the sequence and channel dimensions respectively, $\text{Norm}$ is a normalization layer, $\text{FFN}$ is the feed-forward network, and $\mathsf{T}$ denotes the transpose operation.

The resulting feature forms a sequence with the same shape as $X_S$, where each token is a $D_S$-dimensional vector. We refer to these as spectrum tokens.

\subsubsection{2. How can we enable finer condition injection in DiT?} DiT uses adaLN for time and condition injection. While this strategy is effective for class conditioning, it forbids direct interaction between image tokens and spectrum tokens. In \method, our goal is to enhance the model's ability to capture the relationship between the spectrum and the corresponding material design. Therefore, we seek token-level interaction between tokens from these two modalities.

We first project the spectrum tokens to the hidden size of DiT using a lightweight projector, then the spectrum tokens and image tokens are concatenated along the sequence dimension, establishing an in-context conditioning paradigm. Let $\mathbf{X_S} \in \mathbb{R}^{N_s \times D_T}$ denote the embedded spectrum tokens, and 
$\mathbf{X_I} \in \mathbb{R}^{N_i \times D_T}$ denote the embedded image tokens, where 
$N_s$ and $N_i$ are the number of tokens for the spectrum and image, respectively, and 
$D_T$ is the embedding dimension of DiT. The concatenated self-attention is
\begin{equation}
    \mathbf{X}_{\text{attn}} = \mathrm{Attn}\left([\mathbf{X_I}; \mathbf{X_S}] \in \mathbb{R}^{(N_i + N_s) \times D_T}\right)
\end{equation}
we then discard the spectrum portion and retain only the updated image tokens:
\begin{equation}
    \mathbf{X_I}' = \mathbf{X}_{\text{attn}}[{:}N_i] \in \mathbb{R}^{N_i \times D_T}
\end{equation}
This design enables the model to jointly attend to both modalities and dynamically integrate contextual information into each token representation through self-attention. Notably, it introduces no additional parameters compared to explicitly adding a cross-attention module. Furthermore, we empirically show that this in-context self-attention mechanism outperforms the cross-attention baseline in Section~\ref{sec:ablation}.

We further inject coarse condition control into the model. Let
$\mathbf{t} \in \mathbb{R}^{D_T}$ denote the timestep embedding. We first compute a pooled representation of the spectrum:
\begin{equation}
    \mathbf{s}_{\text{pool}} = \frac{1}{L_s}\sum_{i=1}^{L_s} \mathbf{X_S}[i,:]
\end{equation}
We then combine this with the timestep embedding:
\begin{equation}
    \mathbf{z} = \mathbf{t} + \mathbf{s}_{\text{pool}} \in \mathbb{R}^{D_T}
\end{equation}
the resulting condition signal $\mathbf{z}$ is then used for adaLN modulation. This coarse-to-fine conditioning allows the model to learn at two different levels of granularity.

\subsection{Training Strategy}
We conduct a two-stage training strategy for \method.

\subsubsection{Stage 1: Contrastive Pretraining.} In the first stage, we aim to train the spectrum encoder to learn semantically rich representations by aligning spectral and material features. We adopt a CLIP-style~\cite{radford2021learning} contrastive training paradigm, jointly optimizing the spectrum encoder and a Vision Transformer that encodes the corresponding material design $U$. We extract the representation of $U$ using the \texttt{[CLS]} token from ViT. Let $\mathcal{E}_S$ and $\mathcal{E}_U$ denote the spectrum encoder and the ViT encoder, respectively. The training objective is defined as:
\begin{equation}
\begin{aligned}
&\mathcal{L}_{\text{CLIP}} = \frac{1}{2}\left[\mathrm{CE}(e^\tau\cdot \mathbf{U}^\mathsf{T}\mathbf{S})+\mathrm{CE}(e^\tau\cdot \mathbf{S}^\mathsf{T}\mathbf{U})\right] \\
&\mathbf{U} = \frac{\mathcal{E}_U(U)}{\|\mathcal{E}_U(U)\|_2}\quad \mathbf{S} = \frac{\mathcal{E}_S(S)}{\|\mathcal{E}_S(S)\|_2}
\end{aligned}
\end{equation}
where $\tau$ is a leanbale temperature parameter and $\mathrm{CE}$ represents Cross Entropy loss.

\subsubsection{Stage 2: Diffusion Training.}
In the second stage, we leverage the pre-computed spectral features from \( \mathcal{E}_S \) to train the \method model via a denoising diffusion objective. Given a material structure \( U \), we apply the forward diffusion process to corrupt it at timestep \( t \): $U_t = \sqrt{\bar{\alpha}_t} \, U + \sqrt{1 - \bar{\alpha}_t} \, \epsilon$,
where \( \bar{\alpha}_t = \prod_{i=1}^{t} \alpha_i \), \( \alpha_t \) is the noise schedule, and \( \epsilon \sim \mathcal{N}(0, I) \) is standard Gaussian noise.

The model is trained to predict the added noise, conditioned on the spectrum \( S \), using the following objective:
\begin{equation}
    \mathcal{L}_{\text{diffusion}} = \mathbb{E}_{U, t, \epsilon, S} \left\| \epsilon - \epsilon_\theta(U_t, t, S) \right\|_2^2
\end{equation}
where \( \epsilon_\theta \) denotes the noise prediction model. 

To further enhance conditional generation quality, we adopt classifier-free guidance~\cite{ho2022classifier} during diffusion training and sampling. Specifically, we randomly drop the spectral condition $S$ during training, replacing it with a null embedding. At inference time, model predicts the noise using the following equation$
\hat{\epsilon}_\theta = \epsilon_\theta(U_t, t, \varnothing) + w(\epsilon_\theta(U_t, t, S)-\epsilon_\theta(U_t, t, \varnothing))$,
where $\epsilon_\theta(U_t, t, S)$ is the noise prediction conditioned on the spectrum $S$, 
$\epsilon_\theta(U_t, t, \varnothing)$ is the unconditional prediction, and $w$ is the guidance scale hyperparameter that controls the strength of conditioning. The overview of the proposed architecture and the training strategy are shown in Figure~\ref{fig:main diagram}.

\subsection{Evaluation}
\begin{table}[t]
\small
\centering
\setlength{\tabcolsep}{4mm}{
\begin{tabular}{lcc}
\toprule
Model & \#Param & MAE \\
\midrule
PNN-1~\cite{an2020deep} & - & 0.0539 \\
PNN-2~\cite{Zhangdiffusionmetasurface} & - & 0.0426 \\
\rowcolor[HTML]{EFEFEF} 
\textbf{StarNet-MLP (Ours)} & 1.90M & \textbf{0.0084} \\
\bottomrule
\end{tabular}
}
\caption{\textbf{Prediction error of the surrogate model.} Our model achieves significantly lower spectrum estimation error compared to prior works that employ a dedicated Prediction Neural Network (PNN) for surrogate modeling. Results adopted from the original paper.}
\label{tab:surrogate}
\end{table}


In this paper, we use Accumulated Absolute Error (AAE), defined as $\mathrm{AAE} = \sum_f |S_{\text{gt}}(f)-\hat S(f)|$, where $S_{\text{gt}}(f)$ is the ground truth spectrum. Our objective is to assess model accuracy across the entire frequency spectrum. While a model may perform well at the majority of frequency points, it can still fail at specific frequencies that are critical for the desired functionality. Conventional approaches that average performance across the frequency range can obscure such localized failures, masking important discrepancies that may significantly impact practical applications.

Furthermore, we propose an average AAE metric across $K$ independently generated designs for the same target spectrum, defined as $\text{AAE\&K}=\max_i\{\text{AAE}_i\}_{i=1}^K$. This metric is designed to evaluate the model’s ability to consistently generate multiple distinct yet accurate solutions.

Following~\cite{zhang2024addressing,Zhangdiffusionmetasurface}, we adopt a surrogate model $M_{\text{sur}}$ to predict the spectral response of a given unit cell $U$, using its output $S_{\text{gt}} = M_{\text{sur}}(U)$ as a substitute for computationally expensive electromagnetic simulations. Specifically, we employ StarNet~\cite{ma2024rewrite} equipped with an MLP head to perform the spectrum prediction. As demonstrated in Table~\ref{tab:surrogate}, the surrogate model achieves remarkably low prediction error, validating its effectiveness as a fast and reliable approximation.

\section{Experiment}
In this section, we aim to answer the following questions: (1) Can \method outperform all baselines? (2) Are all components of \method essential? (3) Can we simply scale up \method for better performance? We first elaborate our experimental setups in Section~\ref{sec:expset} and answer the above questions in Section~\ref{sec:mainret}, Section~\ref{sec:ablation} and Section~\ref{sec:scaling}.

\begin{figure*}[t]
    \centering
    \includegraphics[width=0.7\linewidth]{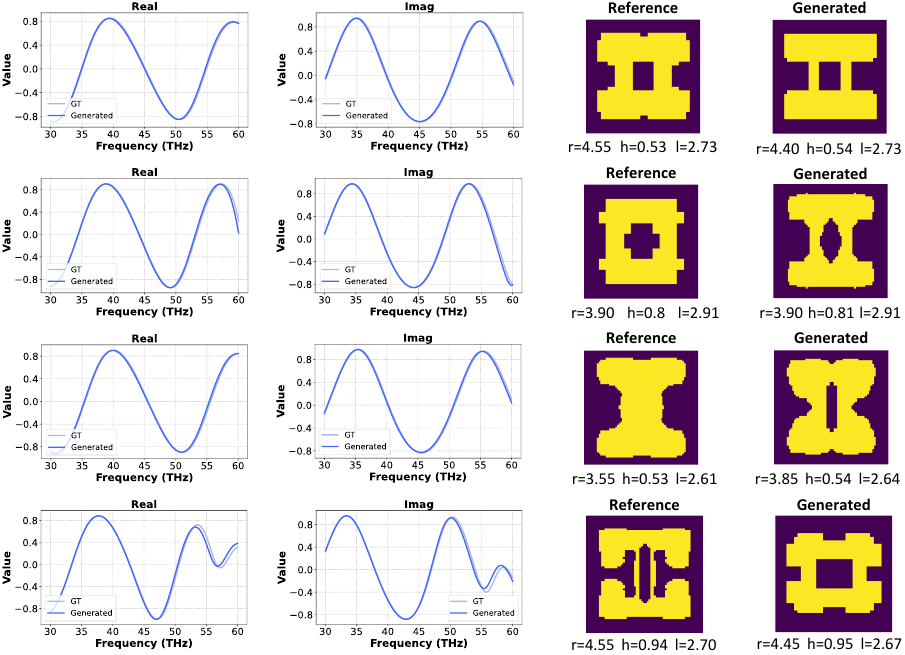}
    \caption{\textbf{Visualization of the generated results.} (Left) Comparsion of the ground truth transmission spectrum and the generated one. (Right) Comparsion bettwen reference material and the generated material. $r$, $h$, $l$ represents $r_{\text{atom}}$, $h_{\text{atom}}$, $l_{\text{lattice}}$ respectively. The unit of $h_{\text{atom}}$ and $l_{\text{lattice}}$ is $\mu$m.}
    \label{fig:visualization}
\end{figure*}

\subsection{Experimental Setups}
\label{sec:expset}
\noindent\textbf{Dataset.} We adopt the dataset from~\cite{an2020deep} and encode it following the procedure detailed in Section~\ref{sec:data encode}. The dataset is randomly split into training, validation, and test sets with a ratio of 8:1:1.

\noindent\textbf{Baselines.} We establish two key baselines for comparison: (1) a standard DiT baseline and (2) a carefully reproduced version of MetaDiff~\cite{Zhangdiffusionmetasurface}, trained under our data encoding framework while maintaining fidelity to the original methodology. We also implement an average predictor that predicts the average of the spectrum values as a basic baseline.

\noindent\textbf{Implementation.} For the spectrum encoder, we stack four layers of spectrum encoder blocks and set the hidden dimension to $D_S = 256$. The FFN the encoder is enhanced with SwiGLU~\cite{shazeer2020glu}. For \method, we vary the model capacity by adjusting the hidden size, the projector is a simple linear layer. Both models employ sinusoidal positional embeddings~\cite{vaswani2017attention}. All experiments are optimized using AdamW~\cite{loshchilov2017decoupled}, with a learning rate of $10^{-4}$ and trained for 500 epochs. We apply cosine annealing as the learning rate schedule. For simplicity and fair comparison, we omit Exponential Moving Average (EMA) and weight decay, although these techniques can further enhance performance. More implementation details can be found in Appendix.

\noindent\textbf{Environment.} We implement the model using PyTorch~\cite{paszke2019pytorch}, with DeepSpeed ZeRO 2~\cite{rajbhandari2020zero} for better training efficiency. We use standard acceleration techniques like bfloat16~\cite{kalamkar2019study} and gradient checkpointing~\cite{herrmann2019optimal}, diffusion training is conducted using full precision. All experiments are conducted using 4$\times$Nvidia A100 80GB. For consistency, we fix the random seed to 0.

\subsection{Main Results}
\label{sec:mainret}
\begin{table}[t]
\small
\centering
\setlength{\tabcolsep}{0.9mm}{
\begin{tabular}{lccccc}
\toprule
Model & \#Param & MAE & AAE & AAE\&2 & AAE\&4 \\ \midrule
AVG$^1$ & - & 0.5860 & 352.7424 & 352.7424 & 352.7424 \\
MetaDiff & 32.56M & 0.1861 & 112.0591 & 170.6253 & 258.9915 \\
MetaDiff-HR$^2$ & 33.41M & 0.1315 & 79.1365 & 100.2521 & 125.4889 \\
DiT & 32.80M & 0.1677 & 100.9437 & 138.0702 & 187.7744 \\
\rowcolor[HTML]{EFEFEF} 
\textbf{\method} & 32.57M & \textbf{0.0801} & \textbf{48.2495} & \textbf{58.8007} & \textbf{68.7275} \\ 
\bottomrule
\end{tabular}
}
\caption{\textbf{Main results of the reproduced baselines and \method}. $^1$ means averge baseline, we calculate the MAE and AAE when model designs the average of the spectrum value. $^2$ means high resolution spectrums are used.}
\label{tab:main}
\end{table}

In this section, we demonstrate that \textit{\method can outperform all baselines}. In our setting, the model is required to design all parameters while adhering to high-resolution spectral constraints. As shown in Table~\ref{tab:main}, directly using high-resolution spectra as input improves the performance of MetaDiff by approximately 29.3\%, underscoring the importance of fine-grained conditioning. However, MetaDiff still struggles to fully capture the relationship between spectrums and material structures. \method further improves the performance by an additional \textbf{39.1\%} over MetaDiff-HR.

Following MetaDiff, we also implement a vanilla DiT baseline conditioned on high-resolution spectra represented as a single feature vector. As shown in Table~\ref{tab:main}, this model performs even worse than MetaDiff-HR, highlighting the necessity of more expressive condition injection mechanisms. By incorporating the techniques described in Section~\ref{sec:method}, \method achieves a substantial improvement of \textbf{52.2\%} over the vanilla DiT baseline.

To evaluate \text{AAE\&K}, we sample multiple designs by varying the random seed across $\{0, 7, 42, 3407\}$. As shown in Table~\ref{tab:main}, \method demonstrates greater robustness in generating diverse metasurface designs. Even under the worst-case scenario, which is comparing the maximum AAE across four different samples, \method achieves substantially lower error than all baselines. We visualize several generated results in Figure~\ref{fig:visualization}.

\subsection{Ablation Studies}
\label{sec:ablation}
\begin{table}[t]
\centering
\small
\setlength{\tabcolsep}{5mm}{
\begin{tabular}{lcc}
\toprule
Method & MAE & AAE \\
\midrule
\rowcolor[HTML]{EFEFEF} 
MetaDiT & 0.0801 & 48.2495 \\
w/o Pretrained Encoder & 0.1370 & 82.4838 \\
w/o Coarse condition & 0.0996 & 59.9662 \\
w/ Cross-attention & 0.0927 & 55.8119 \\
\bottomrule
\end{tabular}
}
\caption{\textbf{Ablation results of \method.} Encoded spectrum features significantly improves the performance, In-context condition and coarse-to-fine conditioning are also essential for \method.}
\label{tab:ablation}
\end{table}

In this section, we ablate key components of \method and provide insights into the architectural choices and training strategies that contribute to its performance.

\subsubsection{How important is proper spectrum encoding?}
We investigate the impact of spectrum encoding by replacing the pretrained encoder and projector with a simple MLP projector and retraining the model. As shown in Table~\ref{tab:ablation}, this leads to a substantial performance drop of 41.5\%, highlighting the critical role of the pretrained encoder and sequential spectrum representation. Notably, the performance remains 18.3\% higher than the vanilla DiT baseline, further underscoring the benefits of sequential formatting and the interaction between spectrum and image tokens.

\subsubsection{Is in-context conditioning effective?}
To assess the effectiveness of in-context conditioning, we remove it and instead introduce an additional cross-attention layer after the self-attention on image tokens. In this setup, spectrum tokens serve as keys and values, while image tokens act as queries. This modification increases the parameter count by approximately 21.7\%. However, as shown in Table~\ref{tab:ablation}, this approach underperforms compared to in-context conditioning, validating the effectiveness of our design for enabling direct interaction between image and spectrum tokens.

\subsubsection{Is coarse condition injection necessary?}
While the in-context conditioning mechanism enables fine-grained, token-level interaction between image and spectrum tokens, we further introduce a coarse condition by pooling the spectrum and injecting it via the adaLN modulation. To assess its impact, we ablate the coarse condition by removing it from the adaLN inputs, leaving only the timestep embedding. As shown in Table~\ref{tab:ablation}, this results in a performance drop of 10.6\% compared to \method, validating our coarse-to-fine design. This demonstrates that learning across two levels of granularity enhances model performance.

\subsection{Is simple scaling effective?}
\label{sec:scaling}
\begin{table}[t]
\centering
\small
\setlength{\tabcolsep}{3mm}{
\begin{tabular}{lcccc}
\toprule
Model & \#Param & Width & \#Layer & \#Head \\
\midrule
MetaDiT & 32.57M & 384 & 12 & 6 \\
MetaDiT-B & 57.78M & 512 & 12 & 8 \\
MetaDiT-L & 129.73M & 768 & 12 & 12 \\
\bottomrule
\end{tabular}
}
\caption{\textbf{Model Specifications of \method}, we vary the width and the number of attention heads to implement different sizes of \method.}
\label{tab:modelspec}
\end{table}

\begin{figure}[t]
    \centering
    \includegraphics[width=0.49\linewidth]{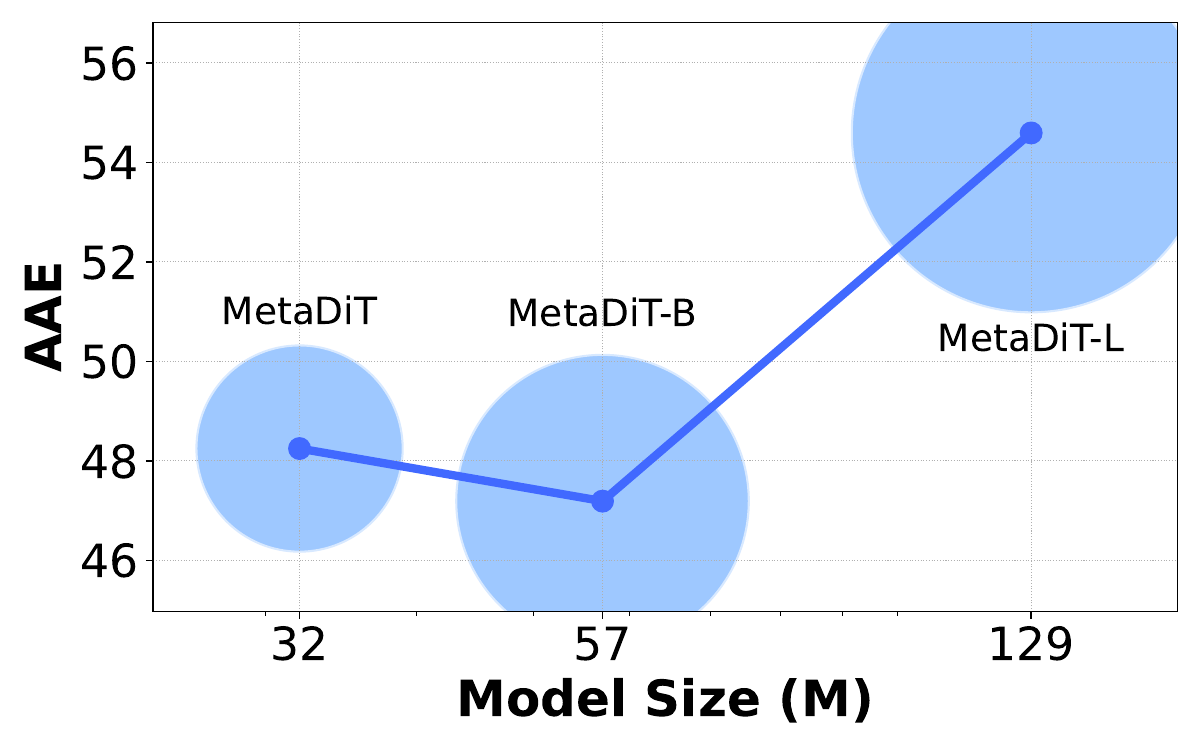}
    \includegraphics[width=0.49\linewidth]{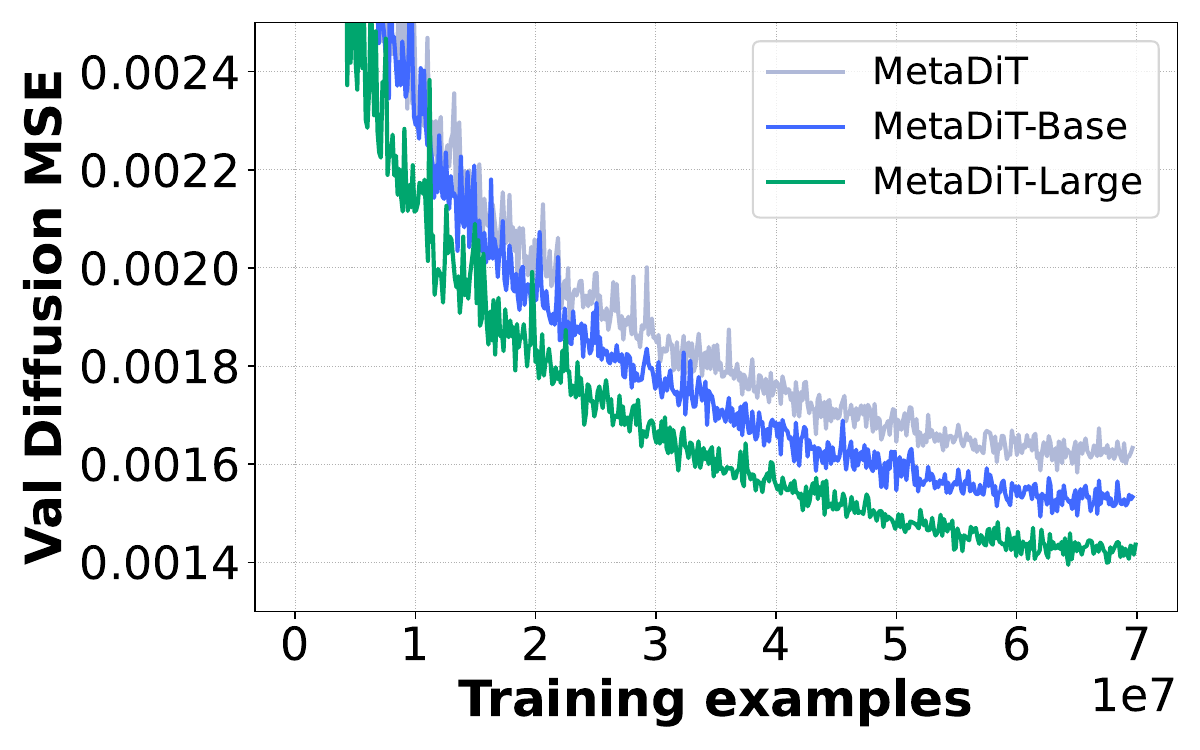}
    \caption{\textbf{Scaling the \method size.} (Left) The final AAE of \method in different sizes. (Right) The MSE loss of model predicted noise on validation set.}
    \label{fig:scaling backbone}
\end{figure}

In this section, we aim to verify that whether better performance can be obtained by simply scaling up the size of the diffusion backbone. Specifically, we adjust the model width (hidden size), number of layers, and number of attention heads. As summarized in Table~\ref{tab:modelspec}, we construct two additional variants of \method. All models are trained under identical settings and evaluated consistently to assess the impact of scale on performance.

The results are presented in Figure~\ref{fig:scaling backbone} (Left). Scaling up to \method-B, which introduces a 77.4\% increase in parameters over \method, yields a modest performance gain of 2.3\%. However, further scaling to \method-L, with an additional 124.5\% increase in parameters relative to \method-B, results in a performance drop of 15.7\%. These findings suggest that simply enlarging the diffusion backbone offers diminishing returns and may even degrade design accuracy.

As illustrated in Figure~\ref{fig:scaling backbone} (Right), scaling up the model size consistently lowers the diffusion MSE on the validation set, suggesting improved reconstruction quality. However, faithfully matching the target spectrum \textbf{requires more than reconstruction fidelity}; it demands precise alignment with the input condition. This is analogous to instruction-following in the visual generation community~\cite{ghosh2023geneval}, where success is measured not solely by image quality, but by how well the output adheres to the input prompt. This insight reinforces our central claim: the fundamental challenge lies in designing a model that effectively captures the intricate relationship between material structure and its corresponding scattering spectrum.

\section{Conclusion}
This paper introduces \method, a novel framework capable of designing the complete set of available metasurface material parameters while accurately satisfying high-resolution scattering spectra. By integrating contrastive pretraining, a dual-attention architecture, and a coarse-to-fine condition injection strategy, \method achieves state-of-the-art performance across all evaluated baselines. Through ablation studies, we highlight the contribution of each design choice. Moreover, scaling experiments reveal that superior design accuracy does not solely stem from improved reconstruction fidelity, but rather from the model’s enhanced ability to learn the underlying physical relationship between material structure and its corresponding scattering spectrum.
\section*{Acknowledgements}
This research was supported by Priority 2030 Federal Academic Leadership Program.
\bibliography{aaai2026}

\newpage
\quad 
\newpage
\appendix

\section{Implementation Details}
\label{sec:implem}
\subsection{Details on Data Encoding}
As described in the original dataset paper~\cite{an2020deep}, each material structure is generated via a random algorithm followed by horizontal / vertical flipping and concatenated. As a result, the pattern matrix exhibits inherent symmetry. To eliminate redundant information and reduce input complexity, we retain only the top-left quadrant of the pattern matrix for downstream processing. For a pattern matrix of shape $P\in \mathbb{R}^{64\times 64}$, we conduct
$$
    P' = P[:32, :32]
$$
and use $P'$ for subsequent processing. After generation, the data is recovered by flipping and concatenating $P'$.

To mitigate potential biases that may adversely affect the performance of activation functions, we normalize the input data to the range $[0, 1]$. Following the original specification in~\cite{an2020deep}, we scale the refractive index and lattice constant by dividing them by their respective maximum values within the dataset. The thickness is left unchanged, as its values naturally fall within the $[0, 1]$ range.

The generated design $\hat{U}$ often contains irregular values due to inherent perturbations, necessitating a binarization step to obtain a valid material structure. To extract the parameters designed by \method, we apply the binarization procedure outlined in Algorithm~\ref{alg:bina}.

\begin{algorithm}[t]
\caption{Binarization of \texttt{gen\_structure}}
\label{alg:bina}
\begin{algorithmic}[1]       
\REQUIRE \texttt{gen\_structure} $\in \mathbb{R}^{3\times 64\times 64}$
\FOR{$i \leftarrow 0$ \TO 2}
    \STATE $\mu \leftarrow \mathrm{mean}\!\bigl(\text{gen\_structure}[i]\bigr)$
    \STATE $mask \leftarrow \bigl(\text{gen\_structure}[i] < \mu\bigr)$
    \STATE $\text{gen\_structure}[i][mask] \leftarrow 0$
    \STATE $m_{\max} \leftarrow \operatorname{clip}\!\bigl(\max(\text{gen\_structure}[i]),\,0,\,1\bigr)$
    \STATE $\text{gen\_structure}[i][\neg\,mask] \leftarrow \operatorname{round}\!\bigl(m_{\max},\,2\bigr)$
\ENDFOR
\STATE \textbf{return} \texttt{gen\_structure}
\end{algorithmic}
\end{algorithm}

\subsection{Details on Model Implementation}
\subsubsection{Surrogate Model.} We adopt StarNet-s3~\cite{ma2024rewrite} as the backbone for extracting features from material designs. To better adapt the model for spectral prediction, we replace the original linear probe head with a two-layer MLP: the first layer expands the input feature dimension by a factor of two, while the second layer projects the output to match the number of frequency points. We use ReLU6 as the activation function to improve boundedness during training.

\subsubsection{Spectrum Encoder.} In the Spectrum Encoder, we employ single-head attention for both sequence and channel attention modules. To enhance training stability, we adopt QK-Norm~\cite{henry2020query} during attention computation. The intermediate dimension of the SwiGLU activation is set to three times the hidden size, following best practices for improving model expressiveness.

\subsection{Details on Training and Inference}
\begin{table}[t]
\small
\centering
\setlength{\tabcolsep}{3mm}{
\begin{tabular}{lccc}
\toprule
 & \begin{tabular}[c]{@{}c@{}}Surrogate\\ Model\end{tabular} & \begin{tabular}[c]{@{}c@{}}Spectrum\\ Encoder\end{tabular} & MetaDiT \\ \midrule
Epoch & 500 & 300 & 500 \\
LR & 1e-3 & 2e-5 & 1e-4 \\
Batch size & 2048 & 4096 & 1024 \\
LR Scheduler & \multicolumn{3}{c}{Cosine} \\
Optimizer & \multicolumn{3}{c}{AdamW} \\
Precision & BF16 & BF16 & FP32 \\ \bottomrule
\end{tabular}
}
\caption{\textbf{Hyperparameters for training.} Here, we report the hyperparameters used in training for all models. Reproduced baselines follow the training recipe of \method.}
\label{tab:hyperpar}
\end{table}
The detailed training hyperparameters are listed in Table~\ref{tab:hyperpar}.

During inference, we set the classifier-free guidance (CFG) scale to 4.0. To compute the AAE\&K metric, we sample multiple material designs under different random seeds, specifically $\{0, 7, 42, 3407\}$, as detailed in the main text. This setup allows us to evaluate the model's robustness and diversity across multiple generation attempts.

\section{Reproducibility}
We open-source all code, data, and model weights to the community. The code repository includes detailed comments and usage instructions to facilitate understanding and reproducibility. We make every effort to ensure that our results can be reliably reproduced.

\end{document}